%% file: main.tex
\documentclass{article}

\usepackage{microtype}
\usepackage{graphicx}
\usepackage{booktabs} %
\usepackage{xcolor}
\usepackage{hyperref}
 \usepackage[square, comma, sort&compress, numbers]{natbib}

\usepackage[final]{changes}

\usepackage[accepted]{mlsys2022}

\usepackage{amsmath}  %
\usepackage[frozencache,cachedir=.]{minted}
\usepackage{xspace}  %
\usepackage{xurl} %
\usepackage{float}
\usepackage{enumitem} %

\usepackage{caption}
\usepackage{subcaption}  %
\usepackage[marginal]{footmisc}

\newcommand{\IGNORE}[1]{}

\newcommand{\sectionshrinker}{}

\newcommand{\extrashrink}[1]{}

{\end{list}}

{\end{list}}

\mlsystitlerunning{HETHUB: \replaced{A distributed training system with heterogeneous cluster for large-scale models}{A Heterogeneous distributed hybrid training system for large-scale models}}

\usepackage{multirow}
\begin{document}

\twocolumn[
\mlsystitle{HETHUB: \replaced{A distributed training system with heterogeneous cluster for large-scale models}{A Heterogeneous distributed hybrid training system for large-scale models}}

\mlsyssetsymbol{equal}{*}
\mlsyssetsymbol{lead}{+}

\begin{mlsysauthorlist}
\mlsysauthor{Si Xu}{i,equal}
\mlsysauthor{Zixiao Huang}{t,i,equal}
\mlsysauthor{Yan Zeng}{h,i,equal}
\mlsysauthor{Shengen Yan}{i}
\mlsysauthor{Xuefei Ning}{t}
\mlsysauthor{Quanlu Zhang}{i}
\mlsysauthor{Haolin Ye}{i}
\mlsysauthor{Sipei Gu}{i}
\mlsysauthor{Chunsheng Shui}{i}
\mlsysauthor{Zhezheng Lin}{i}
\mlsysauthor{Hao Zhang}{c}
\mlsysauthor{Sheng Wang}{c}
\mlsysauthor{Guohao Dai}{s,i,lead}
\mlsysauthor{Yu Wang}{t}
\vspace{-0.05in}
\end{mlsysauthorlist}

\mlsysaffiliation{i}{Infinigence-AI}
\mlsysaffiliation{h}{Hangzhou Dianzi University}
\mlsysaffiliation{t}{Tsinghua University}
\mlsysaffiliation{s}{Shanghai Jiao Tong University}
\mlsysaffiliation{c}{China Mobile Research Institute}

\mlsyscorrespondingauthor{Guohao Dai}{daiguohao@sjtu.edu.cn}
\mlsyscorrespondingauthor{Yu Wang}{yu-wang@tsinghua.edu.cn}

\mlsyskeywords{machine learning, LLMs, inference}
\vskip 0.3in

\begin{abstract}
\input{abstract}
\end{abstract}

]

\printAffiliationsAndNotice{\\ \mlsysEqualContribution \mlsysProjectLead}  %

\widowpenalty0
\clubpenalty0

\sectionshrinker

\input{introduction}

\input{background}

\input{design-implementation}

\input{experiments}

\input{conclusion}


\pagebreak

\input{references}
\clearpage
\vfill
\pagebreak
\newpage

\end{document}

%% file: abstract.tex

\replaced{Training large-scale models}{The development of large-scale models} relies on \replaced{ a vast number of computing resources}{large-scale computing power}. For example, \added{training} the GPT-4 model (1.8 trillion parameters) requires 25000 A100 GPUs \deleted{for its training}. It is a challenge to build a large-scale cluster with \replaced{one}{a} type of GPU-accelerator. Using \replaced{multiple types of}{multi-type} GPU-accelerators to construct a \added{large-scale }cluster \deleted{and train large-scale models} is an effective way to solve the problem of insufficient \added{homogeneous} GPU-accelerators. However, the existing distributed training systems for large-scale models only support homogeneous GPU-accelerators, not \added{support }heterogeneous GPU-accelerators. To address the problem, this paper proposes a distributed training system with hybrid parallelism \replaced{, HETHUB, for large-scale models, which supports}{support on} heterogeneous \added{cluster, including AMD, Nvidia GPU and other types of } GPU-accelerators \deleted{for large-scale models}. It introduces a distributed unified communicator to realize the communication between heterogeneous GPU-accelerators, a distributed performance predictor and an automatic parallel planner to develop and train models efficiently with heterogeneous GPU-accelerators. Compared to the distributed training system with homogeneous GPU-accelerators, our system can support six combinations of heterogeneous GPU-accelerators. We train the Llama-140B model on a heterogeneous cluster with 768 GPU-accelerators(128 AMD and 640 GPU-accelerator A). The experiment results show that the optimal performance of our system in the heterogeneous cluster has achieved up to 97.49$\%$ of the theoretical upper bound performance.


%% file: introduction.tex
\section{Introduction}\label{sec:intro}

\global\csname @topnum\endcsname 0
\global\csname @botnum\endcsname 0

With the rapid development of artificial intelligence technology, large language models (LLMs) like GPT-4 \cite{article_1.0_1}, Pangu \cite{article_1.0_2}, M6 \cite{article_1.0_3}, and others \cite{article_1.0_4,article_1.0_5,article_1.0_6} have grown rapidly, and these models are widely used in various fields \cite{article_1.0_7,article_1.0_8} due to their exceptional performance. However, their parameter scales are typically enormous, ranging from millions to billions, trillions, or even quadrillions. \replaced{It needs a large number of GPU-accelerators to train these large-scale models}{This escalation in scale has made the capacity and computing power required for training LLMs prohibitively large}. For example, GPT-4, with approximately 1.8 trillion parameters, requires around 25,000 A100 GPUs \added{for its training}.

At present, researchers \added{mainly }use Megatron-LM \cite{article_1.0_9}, DeepSpeed \cite{article_1.0_10}, or Pytorch with NVIDIA GPU, such as  A100, V100, H800, to train large-scale models. In recent years, many GPU-accelerators from different manufacturers have developed rapidly. However, it is very hard to build a large-scale cluster with a single type of GPU-accelerator in many scenarios. Training the large-scale model with multiple types of GPU-accelerators is an effective way to solve the problem of insufficient \added{homogeneous }GPU-accelerator. For example, the Llama3-70B model with 70 billion parameters needs 900 Nvidia H100 GPUs to train over 10 months. If there are 500 Nvidia H100 GPUs (1000 TOPs), 400 Huawei GPU-accelerators(320 TOPs), and 700 AMD GPU-accelerators (383 TOPs), we can't train the Llama3-70B model with \replaced{a single}{one} type of GPU-accelerators, as neither \added{one type of } GPU-accelerator can meet the demand. But if we use all types of GPU-accelerators, we can train the Llama3-70B model. Thus, \replaced{building a heterogeneous cluster with different types of GPU-accelerators to train large-scale models, is a good way to solve the problem of large-scale computing resources.}{making full use of different types of GPU-accelerators to train large-scale models is a good way to solve the need for large amounts of computing resources. }

The existing distributed training systems \cite{article_1.0_11,article_1.0_12,article_1.0_13} for large-scale models only support homogeneous \replaced{cluster}{GPU-accelerator}. \deleted{In order to train large-scale models with heterogeneous GPU-accelerators, }
\replaced{It is challenging to train large-scale models with heterogeneous clusters due to the differences in architecture and software between different types of GPU-accelerators. }{we analyze heterogeneous GPU-accelerators and their distributed training systems. We find that there are two challenges caused by the different architectures and software environments. }
\textbf{1) Communication challenge.}
\added{Different types of GPU-accelerators cannot communicate directly with each other, as different types of GPU-accelerators have different communication libraries, such as Nvidia GPUs use NCCl, GPU-accelerator C use HCCL.}
\textbf{2) Development and training challenge.} 
\added{It is very difficult to design and implement an optimal distributed training strategy for large-scale models in a heterogeneous cluster. The differences in computation and storage of different types of GPU-accelerators and the computation-communication strong coupling characteristic of large-scale models result in the exponential increase of the number of distributed strategies with the number of heterogeneous GPU-accelerators, layers or operators of models.}
\textbf{3) Accuracy challenge.} 
\added{The accuracy difference of operators on different types of GPU-accelerators will make the accuracy of the model difficult to reach the accuracy of the homogeneous cluster.}


\replaced{This paper focuses on solving the communication development and training challenges, and proposes a distributed training system with heterogeneous clusters for large-scale models.}{In response to the above challenges, we propose a heterogeneous distributed hybrid training system for large-scale models.} Our major contributions can be summarized as follows:

\begin{enumerate}
    \item We construct a distributed unified communicator to support communication between different GPU-accelerators. This communicator includes two communication libraries, one is a CPU-based communicator with Ethernet or IPoIB; the other one is a GPU-based communicator with IB or RoCE, which defines a unified communication interface to adapt to \added{multiple types of }GPU-accelerators.
    \item A distributed performance predictor is proposed to 
    \replaced{help evaluating the distributed training strategy of models on the heterogeneous cluster. We conduct automatic profiling on a small cluster and build the performance evaluation model. Then, this performance evaluation model can be used to make performance predictions to guide the decision of the distributed training strategy on larger-scale clusters.}{assist the development and training of the model with heterogeneous GPU-accelerators. It can automatically profile performance data, build performance evaluation models, make performance predictions, and guide the search for automatic parallelism strategy in large-scale heterogeneous clusters with the profiling performance of training in small heterogeneous clusters.}
    \item We introduce an automatic parallel planner, which can automatically search an optimal distributed parallelism strategy for the given model and heterogeneous cluster topology.  It can enhance development and model computation efficiency.
    \item \added{We verify the performance and scalability of our system HETHUB with the Llama2-140B model in a heterogeneous cluster with 768 GPU-accelerators, which include 128 AMD GPU-accelerators and 640 GPU-accelerators A.}
\end{enumerate}

%% file: background.tex
\section{Background}\label{sec:backg}

\global\csname @topnum\endcsname 0
\global\csname @botnum\endcsname 0

The success of large-scale models in natural language processing, machine translation, and other fields,  stems from their vast number of parameters, which can fully mine the data characteristics. \replaced{Distrinbuted training method is essential for large-scale models, which}{However, the large number of parameters also poses challenges for distributed training of large-scale models.  Distributed training for large-scale models usually} involves data parallelism, tensor parallelism, pipeline parallelism, and auto-parallel strategy. And if we use different types of GPU accelerators to train a large-scale model, it also involves heterogeneous training. Next, we will present these strategies in detail.

\subsection{Data Parallelism}


Data parallelism (DP) \cite{2018Horovod} is a parallelism method with data segmentation. In data parallelism, the dataset is split into multiple sub-datasets, while the model is replicated across GPU-accelerators. Each GPU-accelerator trains its assigned sub-dataset separately and updates model parameters after forward and backward propagation \cite{2020PyTorch}. However, a challenge arises with memory redundancy as each GPU-accelerator stores duplicate copies of model parameters, optimizer states \cite{2011Adaptive}, and gradients. To address this, DeepSpeed \cite{article_1.0_10} introduced the Zero Redundancy Optimizer \cite{DBLP:journals/corr/abs-1910-02054}, distributing model parameters, optimizer states, and gradients across devices. Consequently, during data updates, GPU-accelerators only need to update their respective partitioned parameters instead of the entire model.

\subsection{Tensor Parallelism}


Tensor parallelism (TP) \cite{article_1.0_9} involves simultaneously processing different parts of a tensor across devices. In tensor parallelism, input tensors are partitioned into subtensors and distributed across devices. After completing the computation of a certain layer, a device may transmit partial results to the next layer or other devices for further processing. Thus, communication operations like AllReduce and Broadcast are implemented typically between different layers or within a layer, to facilitate data exchange and synchronization between devices. However, since AllReduce \cite{2018Horovod} requires communication between all nodes and Broadcast needs to transmit data from one node to other nodes, the communication overhead of these two methods is significant with a large number of nodes.


\subsection{Pipeline Parallelism}

Pipeline parallelism (PP) segments the entire model into \replaced{multiple-stages, and schedules these stages to different nodes or GPU-accelerators, where one stage includes $1$ to $k$ model layers. }{groups of consecutive operators, with each forming a pipeline stage allocated to a corresponding GPU-accelerator.} In pipeline parallelism, batches are typically divided into several micro-batches within each pipeline stage, and parameters for each stage are distributed to the respective computational GPU-accelerators during model initialization. This approach optimizes end-to-end training time by overlapping computation and communication across different stages and micro-batches.

Current mainstream pipeline parallelism schemes include Gpipe \cite{article_2.0_1}, PipeDream-1F1B \cite{pipedream-1f1b}, PipeDream-2BW \cite{pipedream-2bw}, and Chimera \cite{article_2.0_3}. Gpipe divides each stage's batch into multiple micro-batches that are computed sequentially, allowing the computation of micro-batches from different stages to overlap. However, in this arrangement, a stage must wait for the forward pass of all micro-batches to complete before commencing the backward pass, necessitating the storage of activations for the number of micro-batches, which results in significant memory consumption. Conversely, the PipeDream-1F1B arrangement interleaves forward and backward computations, \replaced{storing the activation values of $n$ pipeline stages at most}{requiring each pipeline stage to store at most the activations for the number of stages}, thereby reducing memory usage. Although PipeDream-1F1B has a similar pipeline utilization with Gpipe, it introduces a great optimization to the bubble ratio. PipeDream-2BW takes PipeDream-1F1B further by dividing the pipeline into two buckets to improve the overlap of computation and communication. It potentially leads to even higher throughput and better utilization, though at the cost of increased complexity and higher communication load, which are not supported in heterogeneous clusters. Chimera is a bi-directional scheduled pipeline method. Two micro batches start to train at the same time from the first and the last pipeline stage. So it has less pipeline bubble compared to the common one-directional pipeline. However, this method needs each stage to contain two parts of the model parameter, which adds a \added{memory }burden to the GPU-accelerator \deleted{memory}.


\subsection{Auto-Parallel Strategy}
Due to the complexity of model structures, the \replaced{number}{amount} of distributed  \replaced{training}{parallelism} strategies for models increases exponentially with the number of model layers or operators. This makes manual tuning of \replaced{distributed }{parallelism }strategies based on expert experience highly demanding. To address this issue, researchers have proposed automatic parallelism methods. These methods utilize \replaced{dynamic programming or graph algorithms }{techniques such as dynamic programming and graph algorithms} to automatically search for distributed \replaced{training}{parallelism} strategies for models. These approaches can enhance efficiency by automatically identifying optimal or near-optimal distributed \replaced{training}{parallelism} strategies.

In recent years, automatic parallelism algorithms have developed rapidly\cite{automap, fan2021dapple, distir}, \deleted{including notable frameworks} such as FlexFlow \cite{jia2019beyond}, D-rec \cite{article_2.0_5}, Piper\cite{piper}, and Alpa \cite{article_2.0_6}. FlexFlow constructs a SOAP search space and employs the Markov Chain Monte Carlo (MCMC) algorithm to identify the optimal intra-operator parallelism strategy. The Double Recursive Algorithm (D-rec) considers both intra-operator and inter-operator communication costs, utilizing a two-layer recursive algorithm to determine the optimal tensor sharding strategy. 
\replaced{These }{However, these} methods can only search for partial parallelism strategies. For training large-scale models, it is essential to combine multiple parallelism methods. Piper introduces two-level dynamic programming to search for the distributed parallel strategy for data parallelism and model parallelism. Alpa supports the joint search of multiple parallelism strategies, divides the strategies into intra-operator parallelism and inter-operator parallelism, and optimizes intra-operator parallelism using integer linear programming while arranging inter-operator parallelism using dynamic programming. However, \replaced{these methods only support training large-scale models under homogeneous clusters, not support or work well in heterogeneous clusters.}{when GPU-accelerators are heterogeneous, these frameworks' applicability is limited to the assumption that the devices have the same computation time.} 

\subsection{Heterogeneous Training}

Unlike \replaced{distributed}{parallel} training \added{for large-scale models} in a homogeneous cluster with one type of GPU-accelerators, \replaced{it }{the heterogeneous scenario} introduces communication and computational challenges \added{ in a heterogeneous cluster with multiple types of GPU-accelerators. For communication, it impedes the transfer of data between different types of GPU-accelerators, as different types of GPU-accelerators have their own communication libraries, which are not compatible with each other. For computation, existing distributed training strategies can lead to load imbalance, due to the balanced partitioning of tasks and the differences in computing resources of different types of GPU-accelerators.} 



Existing frameworks that support heterogeneous clusters include BPT-CNN \cite{bpt_cnn_2018}, AccPar \cite{accpar_2020}, and Whale \cite{whale_2022}. BPT-CNN, a data parallelism algorithm for heterogeneous \replaced{clusters}{environments}, allocates \added{setting} the number of batches based on \replaced{the computing resources of GPU-accelerators}{each hardware’s computational capacity}. It dynamically adjusts batch distribution after each training iteration according to the actual performance of \replaced{GPU-acclerators}{the hardware}. AccPar, a tensor parallelism algorithm, employs dynamic programming to derive optimal tensor-sharding strategies for each operator. \replaced{These}{However, these parallelism} methods are limited to single parallelism strategy \deleted{configurations}. Whale \replaced{supports both }{extends these capabilities by supporting} data parallelism and pipeline parallelism \deleted{consequently}. It employs data parallelism partition according to \added{computing resources of }GPU-accelerators\deleted{' computational power} and arranges pipeline parallelism stages \replaced{with memory resources of GPU-accelerators}{by GPU memory capacity} in descending order.  In conclusion, compared to training a model in a homogeneous cluster,  \replaced{it needs to solve}{the heterogeneous training needs to solve} two major problems \added{in a heterogeneous cluster}, the communication between different types of GPU-accelerators , and the efficiency of development and training caused by the \replaced{complexity }{composability} of model structure and the performance differences between multiple types of GPU-accelerators.

%% file: design-implementation.tex
\begin{figure*}[!htbp]  
    \centering
    \includegraphics[width=135mm]{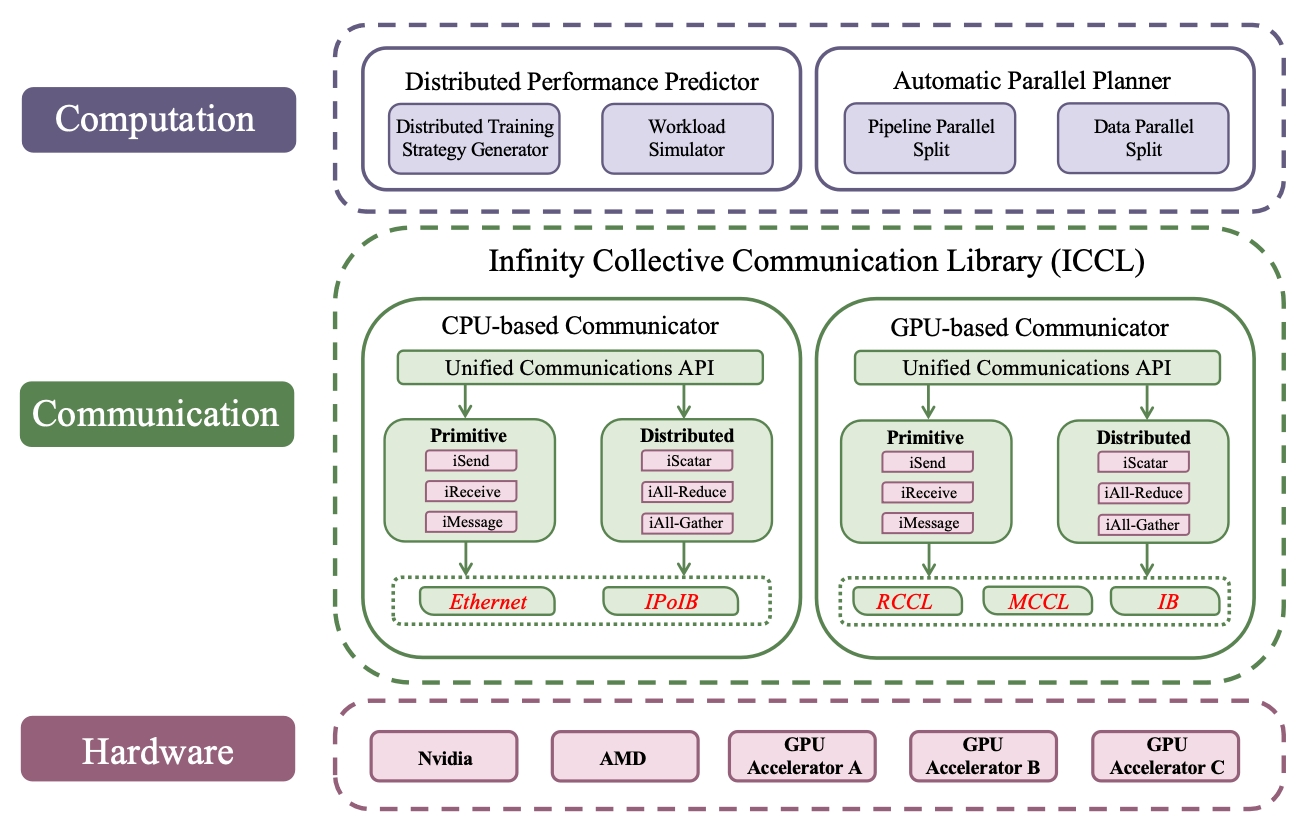}
     \caption{The scheme of HETHUB system}
    \label{1_design_graph}	
\end{figure*}

\section{Design and Implementation}\label{sec:des-imp}

\replaced{HETHUB, a distributed training system }{We propose a heterogeneous distributed hybrid training system} based on Megatron-LM and Megatron-DeepSpeed \cite{article_3.0_2}, for large-scale models in a cluster with different types of GPU-accelerators. \added{We make optimization in the computing and communication level based on the differences of heterogeneous hardware, compared with the existing distributed training system in homogeneous clusters. As shown in Fig. \ref{1_design_graph}, we design an infinity collective communication library(ICCL) based on Gloo \cite{article_3.0_1} to support the communication between heterogeneous GPU-accelerators. It includes a CPU-based communicator with Ethernet or IPoIB \cite{article_3.0_8} and a GPU-based communicator with IB. At the communication level, we propose a distributed performance predictor to predict the training performance of large-scale models, and an automatic parallel planner to search for efficient distributed training strategies in heterogeneous clusters for large-scale models.}
\begin{figure}[!htbp]  
    \includegraphics[width=80mm]{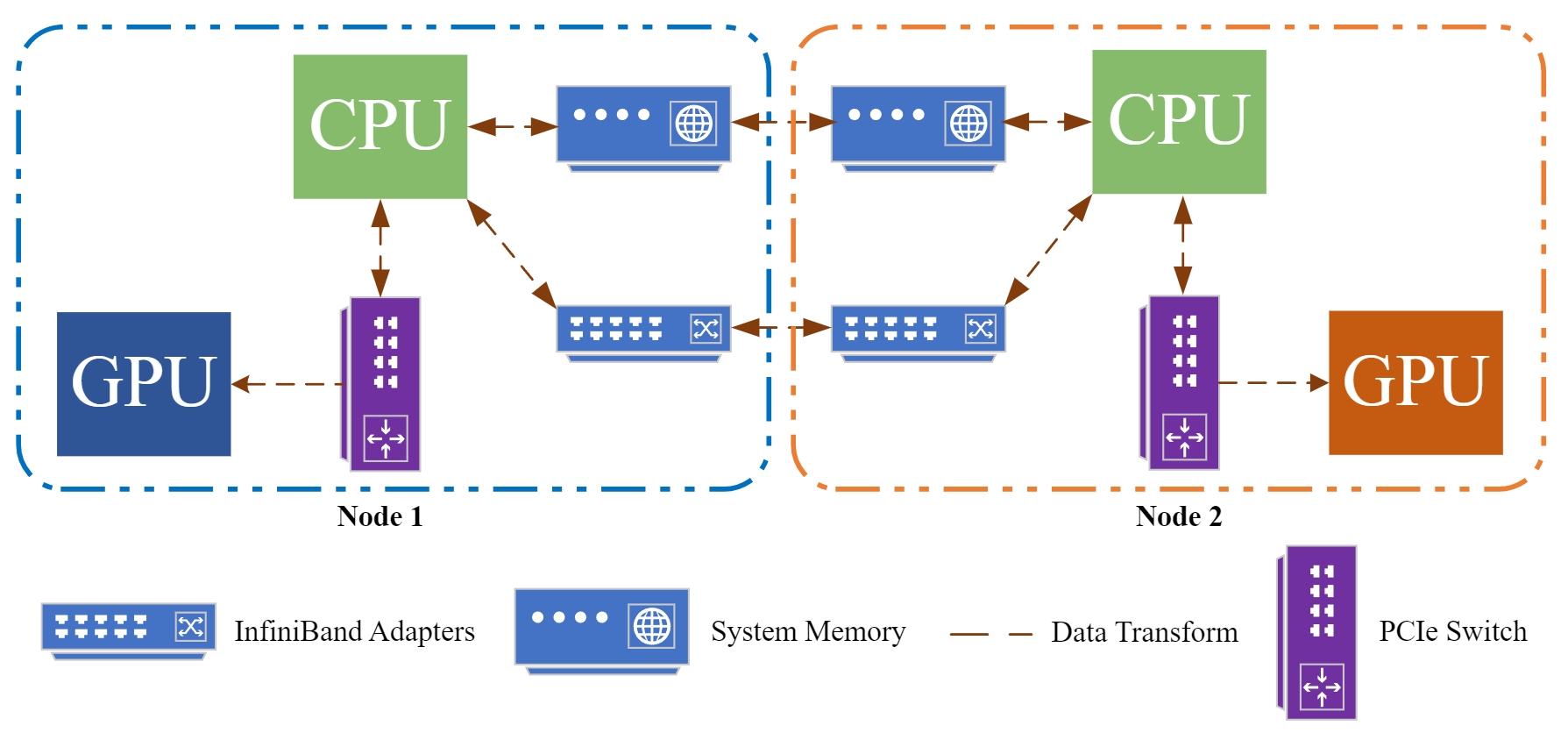}
     \caption{CPU-based communicator with Ethernet or IPoIB}
    \label{2_CPU_communicate}	
\end{figure}


\deleted{Next, we will describe these optimizations in detail.}

\subsection{\replaced{Infinity Collective Communication Library}{Distributed Unified Communicator}}

\added{ICCL, including CPU-based communicator and GPU-based communicator, is proposed to solve the communication problem between different types of GPU-accelerators. It is implemented with Gloo, and supports two types of communication modes, CPU-based communication and GPU-based communication.}

\begin{figure}[!htbp]  
    \centering
    \includegraphics[width=80mm]{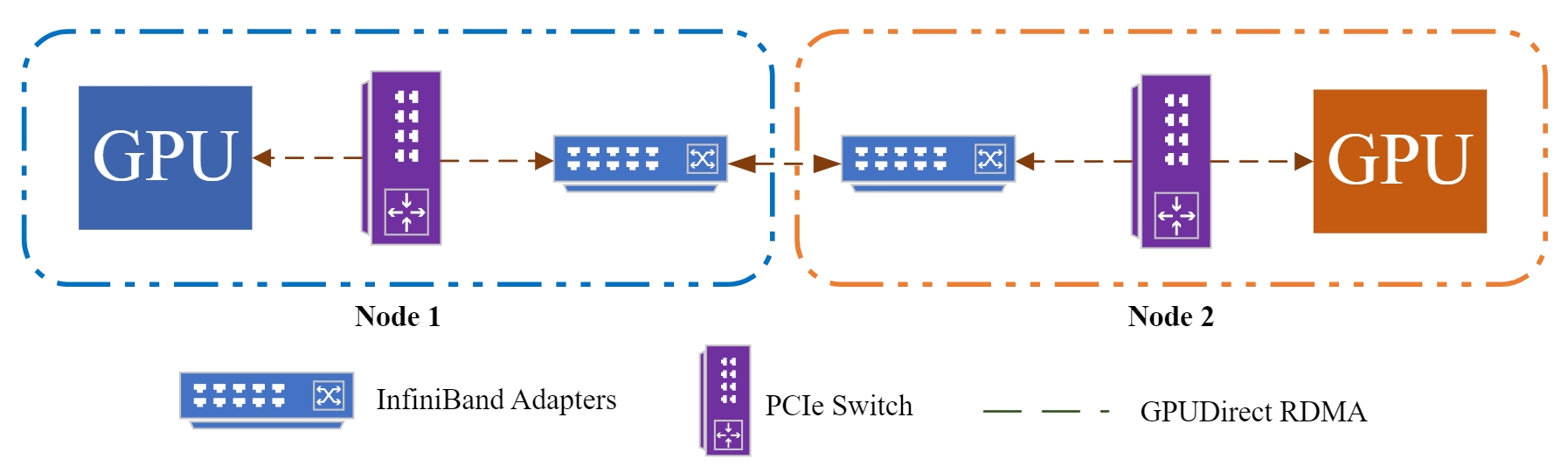}
     \caption{GPU-based communicator with IB}
    \label{3_GPU_communicate}	
\end{figure}

\begin{figure*}[!htbp]  
    \centering
    \includegraphics[width=145  mm]{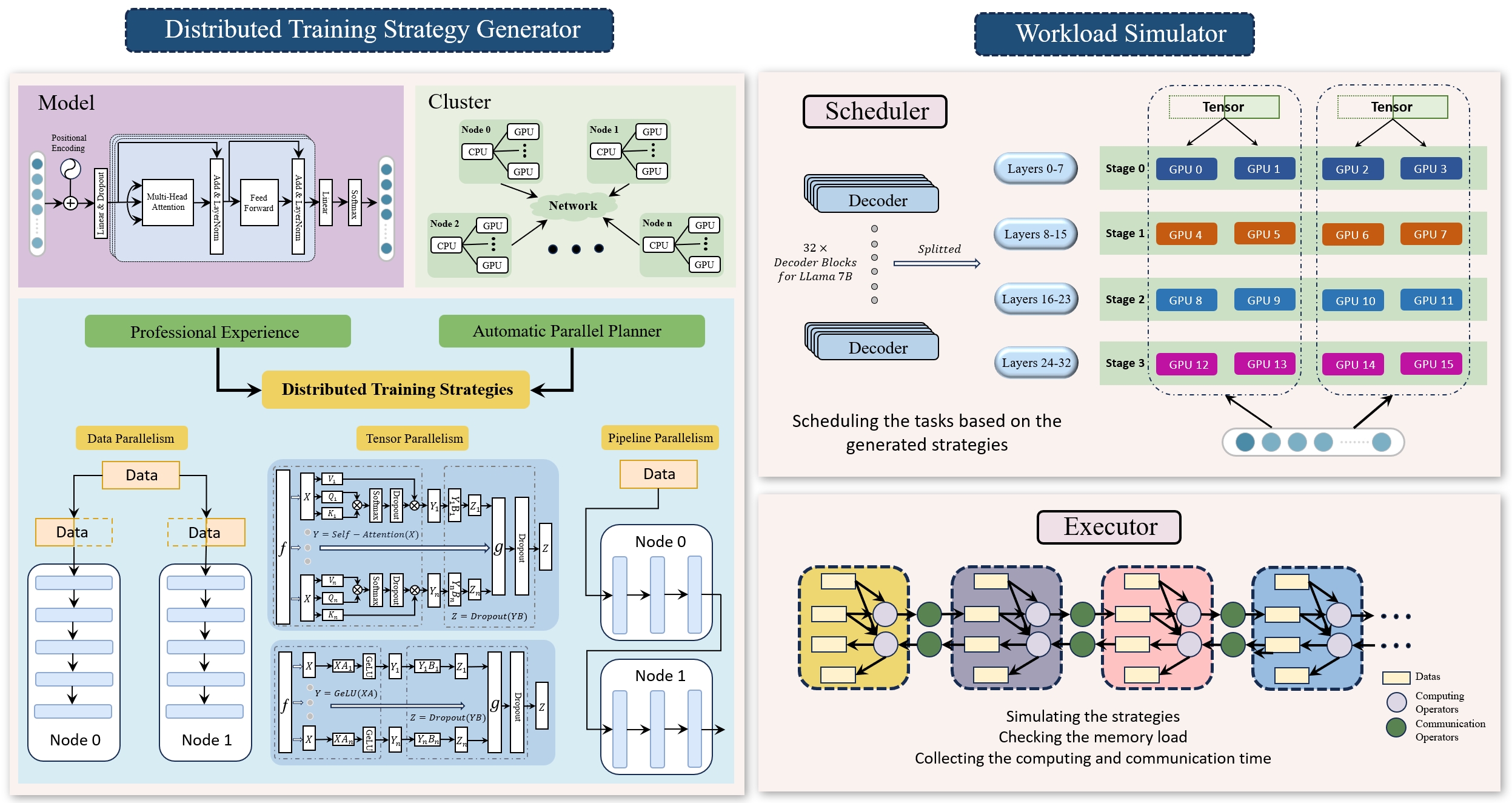}
    \caption{The workflow of distributed performance predictor}
    \label{4_per_predictor}	
\end{figure*}

\textbf{CPU-based Communicator.} This communicator is designed for the scalability of heterogeneous clusters for different types of GPU-accelerator, it supports a new type of GPU-accelerator to join heterogeneous clusters quickly and at a low cost for training large-scale models. Physically, the CPU and the GPU-accelerator are connected through PCIE. CPUs and GPU-accelerators in a node directly communicate with each other via PCIE, and CPUs on different nodes communicate with each other with IPoIB or Ethernet. With the CPU-based communicator, the data, such as model parameters, needs to be copied from the GPU-accelerator to the CPU with PCIE, and then transferred to the CPU of target nodes with Ethernet or IPoIB, and lastly copied the data from the CPU to the GPU-accelerators on the target nodes with PCIE, as shown in Fig. \ref{2_CPU_communicate}. This method avoids the differences in communication libraries of different GPU-accelerators by increasing the copy overhead between CPU and GPU-accelerators.



\textbf{GPU-based Communicatior.} \added{This communicator is designed based on the RDMA \cite{article_3.0_9} for direct communication between heterogeneous GPU-accelerators via IB. Communication is a bottleneck problem in large-scale model training, because of the strong coupling between computation and communication of large-scale models. Different manufacturers have developed special communication libraries for GPU-accelerators to improve communication performance. However, the differences between communication libraries, such as APIs, data types, etc., cause different types of GPU-accelerators to not communicate directly with each other. We define a unified set of distributed communication protocols for different types of GPU-accelerators. It includes data type, communication primitive(e.g. iSend/iRecive), distributed communication function(e.g. iAllReduce, iAll-to-All, and etc). This method has high communication efficiency, but it needs different manufacturers of GPU-accelerators to adapt the unified communication protocol.}



\subsection{Distributed Performance Predictor}
\added{The distributed performance predictor is designed to find distributed training strategies for large-scale models in heterogeneous clusters with low cost. It proposes a distributed training strategy generator to construct different strategies for large-scale models, and presents a workload simulator to simulate and train models with generated distributed training strategy. The optimal distributed training strategy searched by the distributed performance predictor will be used to train large-scale models in real heterogeneous clusters.}

\textbf{Distrbiuted Training Strategy Generator.}
\added{ It constructs a computed graph with weights to describe the structure of large-scale models, where the weights represent the computation performance of the operator or layer of transformer models, sampled from different types of GPU-accelerators. It also samples the computing(e.g. TFLOPs in fp16 precision) and storage information of different GPU-accelerators, as well as the communication bandwidth between heterogeneous GPU-accelerators. Based on this, we use the expert experience or automatic parallel method to generate distributed training strategies for large-scale models, including pipeline parallelism, data parallelism, and tensor parallelism.}

\begin{figure*}[!htbp]  
    \centering
    \includegraphics[width=150mm]{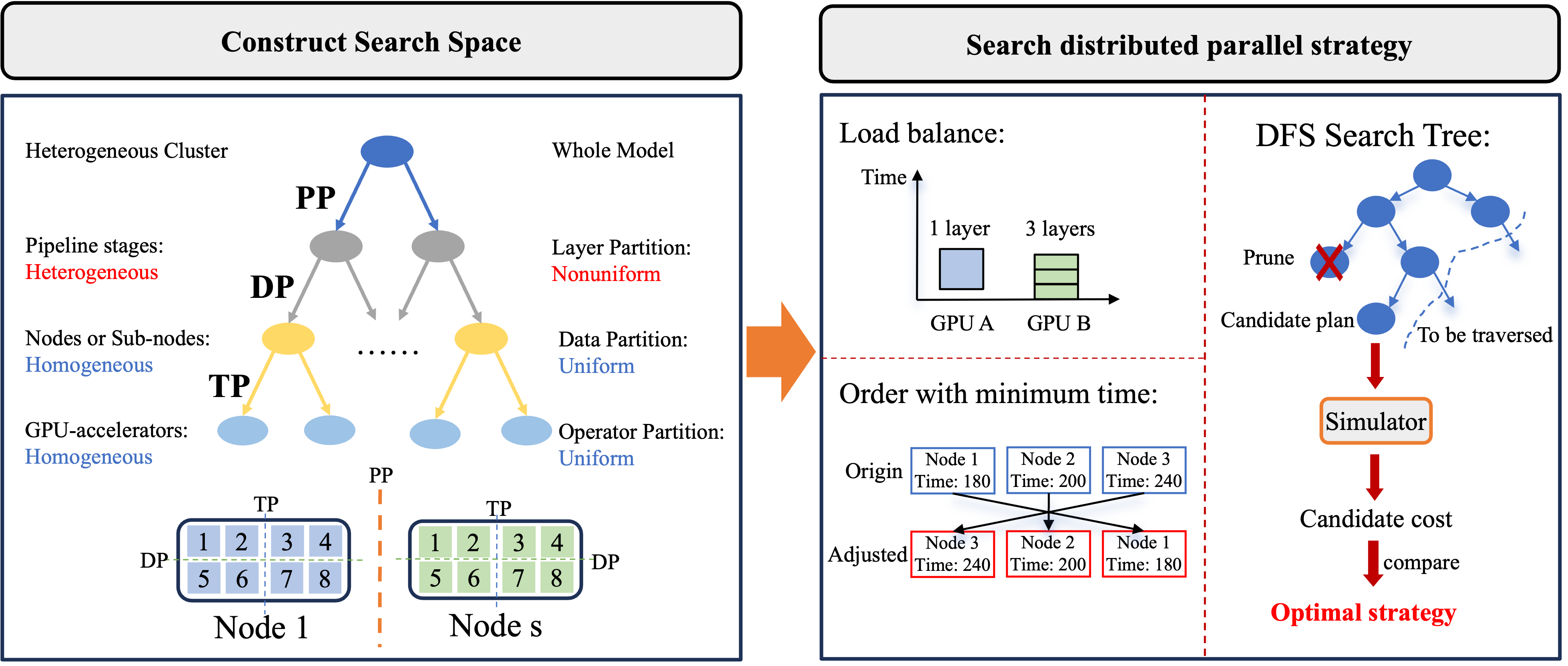}
     \caption{The workflow of automatic parallel planner}
    \label{5_auto_parallel}	
\end{figure*}

\textbf{Workload Simulator.}
\added{The workload simulator is developed to simulate the execution of model training tasks in real heterogeneous clusters. It will calculate the overhead of training the large-scale model, such as the time cost for each iteration step and max memory usage, according to the distributed training strategy generated by the distributed training strategy generator and the operator's execution performance on different types of GPU-accelerators. Lastly, it will give the optimal distributed parallel strategy for a large-scale model from multiple strategies generated by the distributed training strategy generator, to guide the deployment and training of the model in a real heterogeneous cluster.}



\subsection{\replaced{Automatic Parallel Planner}{Automatic Hybrid Parallelism}}

\added{It is an NP-hard problem to find an optimal distributed training strategy for a large-scale model\cite{3_auto_survey, 3_np_hard}, as the number of distributed training strategies increases exponentially with the increase of the number of model layers or operators, especially in heterogeneous clusters. The goal of the automatic parallel planner is to find an optimal distributed training strategy for the model automatically. It automatically divides the large-scale model into multiple sub-models and schedules them to different GPU-accelerators, according to the execution performance of the model layer or operator on different GPU-accelerators. It supports data parallelism, tensor parallelism, and pipeline parallelism for models.} 

\added{Considering the communication performance between heterogeneous GPU-accelerators is lower than that between homogeneous GPU-accelerators, we employ data parallelism combined with intra-node tensor parallelism on homogeneous nodes and pipeline parallelism across heterogeneous nodes. To ensure the correctness of training results, pipeline parallelism satisfies the data constraints of the Pipedream-1F1B scheme. The specific design is illustrated in the Fig. \ref{5_auto_parallel}.}

\added{\textbf{Construct search space.} }
A three-level search tree is constructed to represent the search space of distributed training strategies for models. Where the root node represents the whole model, and the other nodes represent the sub-models after splitting. Furthermore, the leaf nodes represent the final models that are executed in a single GPU-accelerator. The first layer uses a non-uniform pipeline parallelism splitting strategy to split the model based on the total number of transformer layers. The purpose of the splitting is to ensure load balancing of the computation for different types of GPU-accelerator. The second layer splits the sub-model into homogeneous nodes using a uniform data parallelism strategy. The third layer splits the model into GPU-accelerators using a uniform tensor parallelism strategy. After three layers of splitting, the complete model can be mapped to a heterogeneous cluster for training.

\textbf{Search distributed training strategy.} To make full use of heterogeneous GPU-accelerator resources, we give two rules with the goal of load balancing and minimum end-to-end training time, to guide distributed parallelism strategy searching in the constructed search tree.

    {1) Load balance}.   
    According to the computing resources of the heterogeneous GPU-accelerator and the computing requirements of the model layer, we divide the model layer irregularly to balance the computing tasks among different GPU-accelerators as much as possible. That is, GPU-accelerators with high computing resources perform more layers.
    
    \replaced{2) Minimum end-to-end training time.}{Stage order adjustment} 
    \added{We schedule stages in pipeline parallelism to different types of GPU-accelerators to optimize the end-to-end training time, according to the execution time of stages on different types of GPU-accelerators and the communication time between the stages.}

\added{Based on the above rules, we find a distributed training strategy by using the DFS algorithm to traverse the search tree, and use the distributed performance predictor to evaluate the training time of the model with the searched distributed training strategy. Ultimately, we select the distributed training strategy with the lowest evaluation cost.}

%% file: experiments.tex
\section{Experiments}\label{sec:exper}

In order to verify the effectiveness and performance of the \replaced{HETHUB system}{hybrid distributed training system}, this paper uses the models of Llama2 to carry out heterogeneous training experiments on NV, AMD, and other types of GPU-accelerators. In addition, this paper compares the throughput and MFU performance with the homogeneous cluster environment, and also analyses the end-to-end time of model training on a heterogeneous cluster with 768 GPU-accelerators.

\subsection{Experimental Setup}

\added{\textbf{Homogeneous clusters.} We use 2 representative homogeneous clusters, including the AMD GPU-accelerator and GPU-accelerator A clusters. The AMD cluster includes 20 worker nodes with 160 GPU-accelerators, each worker node has 8 GPU-accelerators connected to 192 CPU cores via PCIe switches. The GPU-accelerator A cluster includes 96 worker nodes with 768 GPU-accelerators, each worker node has 8 GPU-accelerators connected to 128 CPU cores via PCIe switches. And infiniband (IB) is used for homogeneous clusters, where the bandwidth is 200 Gb/s.}

\added{\textbf{Heterogeneous cluster.} We use 4 heterogeneous clusters with the ratio of AMD and GPU A being 1:5, including a 12-nodes cluster with 96 GPU-acclerators(12N96D), a 24-nodes cluster with 192 GPU-accelerators(24N192D), a 48-nodes cluster with 384 GPU-accleraotrs(48N384D) and a 96-nodes cluster with 768 GPU-accelerators(96N768D). Infiniband (IB) is used for homogeneous nodes, where the bandwidth is 200 Gb/s. Ethernet is used for heterogeneous nodes, where bandwidth is 25 Gb/s.}

\added{\textbf{Experiment parameters.} We train Llma2 models with the hybrid distributed parallelism strategy with pipeline parallelism, data parallelism, and tensor parallelism. For the heterogeneous cluster, we employ data parallelism combined with intra-node tensor parallelism on homogeneous nodes and pipeline parallelism across heterogeneous nodes. The details of the configurations are shown in Table \ref{tabel1}. Where $DP$ represents data parallelism, $TP$ represents tensor parallelism, $PP$ represents pipeline parallelism, $NUM$ is the number of nodes, global-batch-size=2048*$NUM$/10 for homogeneous clusters, and global-batch-size=2048*$NUM$/6 for heterogeneous clusters, $DP$=$NUM$*8/TP/PP and TP=1 for all clusters, and the ratio of heterogeneous clusters is AMD:GPU A = 1:5.}

\begin{table*} [!h]
 \centering
 \caption{Configuration of model training in homogeneous and heterogeneous clusters.  }
 \begin{subtable}[h]{0.495\linewidth}
        \begin{tabular}{ccccc}
            \toprule
            model & Layers & Hidden Size & PP & NUM  \\
            \midrule
            \multirow{4}*{Llama2-7B} & \multirow{4}*{32} & \multirow{4}*{4096} & \multirow{4}*{12} & 12 \\
            ~ & ~ & ~ & ~ & 24 \\
            ~ & ~ & ~ & ~ & 48 \\
            ~ & ~ & ~ & ~ & 96 \\
            \bottomrule
        \end{tabular}
        \captionsetup{font={small,bf}, justification=raggedright}
        \caption{Configurations of Llama2-7B in heterogeneous clusters}
    \end{subtable}
    \vspace{3em}
    \begin{subtable}[h]{0.495\linewidth}
        \begin{tabular}{ccccc}
            \toprule
            model & Layers & Hidden Size & PP & NUM  \\
            \midrule
            \multirow{4}*{Llama2-13B} & \multirow{4}*{40} & \multirow{4}*{5120} & \multirow{4}*{12} & 12 \\
            ~ & ~ & ~ & ~ & 24 \\
            ~ & ~ & ~ & ~ & 48 \\
            ~ & ~ & ~ & ~ & 96 \\
            \bottomrule
        \end{tabular}
        \captionsetup{font={small,bf}, justification=raggedright}
        \caption{Configurations of Llama2-13B in heterogeneous clusters}
    \end{subtable}
    \vspace{3em}
    \begin{subtable}[h]{0.495\linewidth}
        \begin{tabular}{ccccc}
            \toprule
            model & Layers & Hidden Size & PP & NUM  \\
            \midrule
            \multirow{4}*{Llama2-35B} & \multirow{4}*{40} & \multirow{4}*{8192} & \multirow{4}*{12} & 12 \\
            ~ & ~ & ~ & ~ & 24 \\
            ~ & ~ & ~ & ~ & 48 \\
            ~ & ~ & ~ & ~ & 96 \\
            \bottomrule
        \end{tabular}
        \captionsetup{font={small,bf}, justification=raggedright}
        \caption{Configurations of Llama2-35B in heterogeneous clusters}
    \end{subtable}
    \begin{subtable}[h]{0.495\linewidth}
        \begin{tabular}{ccccc}
            \toprule
            model & Layers & Hidden Size & PP & NUM  \\
            \midrule
            \multirow{4}*{Llama2-70B} & \multirow{4}*{80} & \multirow{4}*{8192} & \multirow{4}*{12} & 12 \\
            ~ & ~ & ~ & ~ & 24 \\
            ~ & ~ & ~ & ~ & 48 \\
            ~ & ~ & ~ & ~ & 96 \\
            \bottomrule
        \end{tabular}
        \captionsetup{font={small,bf}, justification=raggedright}
        \caption{Configurations of Llama2-70B in heterogeneous clusters}
    \end{subtable}
    \begin{subtable}[h]{0.495\linewidth}
        \begin{tabular}{ccccc}
            \toprule
            model & Layers & Hidden Size & PP & NUM  \\
            \midrule
            \multirow{4}*{Llama2-140B} & \multirow{4}*{160} & \multirow{4}*{8192} & \multirow{4}*{24} & 12 \\
            ~ & ~ & ~ & ~ & 24 \\
            ~ & ~ & ~ & ~ & 48 \\
            ~ & ~ & ~ & ~ & 96 \\
            \bottomrule
        \end{tabular}
        \captionsetup{font={small,bf}, justification=raggedright}
        \caption{Configurations of Llama2-140B in heterogeneous clusters}
    \end{subtable}
    \begin{subtable}[h]{0.495\linewidth}
        \begin{tabular}{cccccc}
            \toprule
            model & Layers & Hidden Size & PP & \multicolumn{2}{c}{cluster\hspace{1em}NUM}  \\
            \midrule
            \multirow{5}*{Llama2-70B} & \multirow{5}*{80} & \multirow{5}*{8192} & \multirow{5}*{10} & \multirow{2}{*}{AMD} & 10 \\
            ~ & ~ & ~ & ~ & ~ & 20 \\
            \cline{5-6}
            ~ & ~ & ~ & ~ & \multirow{3}*{GPU A} & 60 \\
            ~ & ~ & ~ & ~ & ~ & 80 \\
            ~ & ~ & ~ & ~ & ~ & 96 \\
            \bottomrule
        \end{tabular}
        \captionsetup{font={small,bf}, justification=raggedright}
        \caption{Configurations of Llama2-70B in homogeneous clusters}
    \end{subtable}
 \label{tabel1}%
\end{table*}

\subsection{Model and Dataset}

\textbf{Model.} We choose the language model Llama2 as the test model. The Llama2 \cite{article_4.0_1}  is a collection of pre-trained and fine-tuned large language models (LLMs) ranging in scale from 7 billion to 140 billion parameters, \added{including Llama2-7B, Llama2-13B, Llama2-35B, Llama2-70B and Llama2-140B}.

\textbf{DataSet.} We conduct experiments on the Dolma and MAP-CC datasets. Dolma Dataset \cite{soldaini2024dolma} is an open dataset of 3 trillion tokens from a diverse mix of web content, academic publications, code, books, and encyclopedic materials. MAP-CC (Massive Appropriate Pretraining Chinese Corpus) dataset \cite{mapcc_2024} is a large-scale open-source Chinese pre-training dataset developed by Multimodal Art Projection, Fudan University, Peking University, and other organizations. \deleted{The dataset is a large-scale open-source Chinese pre-training dataset.} The dataset contains 80 billion tokens and consists of multiple subsets, each from different data sources, such as blogs, news articles, Chinese encyclopedias, Chinese academic papers, and Chinese books.

\subsection{Metrics}

\added{\textbf{Throughput.} Throughput is measured by the number of tokens calculated by one GPU-accelerator in one second. It can be calculated according to Eq. \ref{calc_th}}

\begin{equation}\label{calc_th}
    TGS = \frac{L\times G}{S\times T}
\end{equation}

\added{$L$ represents the sequence length of the training data, $G$ means the global batch size of one iteration, $S$ represents the number of GPU accelerators being used in the training process, and $T$ represents the time spent on training one iteration.}

\added{\textbf{MFU.} Model FLOPs Utilization (MFU) is the ratio of the actual throughput to the theoretical maximum throughput assuming 100\% of peak FLOPs.The definition of MFU is shown in Eq. \ref{MFU_eq}. We calculate the MFU in the same environment as the previous section was used. }
\begin{equation}\label{MFU_eq}
    \begin{aligned}
        \text{MFU} = \frac{T_{test}}{T_{peak}} \times 100\%
    \end{aligned}
\end{equation}
\added{\noindent where $T_{test}$ stands for the tested TFLOPS in fp16 precision per GPU for cluster and $T_{peak}$ refers to the theoretical peak TFLOPS in fp16 precision. For the heterogeneous cluster, the peak TFLOPS is the average value of all GPU-accelerators.}

\added{\textbf{Communication.} Communication is used to measure the cost of communication between different nodes for training models in this paper. The point-to-point communication $T_{com}$ in the model training is defined as Eq.\ref{p2p_com}:}
\begin{equation}\label{p2p_com}
    T_{com} = B \times L \times H \times 2
\end{equation}
\added{\noindent Where the $B$ represents batch size, $L$ is the sequence length and $H$ denotes the number of features in each hidden state.}

\subsection{\replaced{Experiment Results}{Performance Analysis}}

\subsubsection{Throughput}






\added{We first use the Llama2-7B model to verify the effect of uniform and un-uniform stage segmentation of pipeline parallelism on the performance of a small heterogeneous cluster with 1 AMD GPU-accelerators and 5 GPU-accelerator A. Then, we train Llam2-7B, Llama2-13B, Llama2-35B, Llama2-70B and Llama2-140B on different heterogeneous clusters with un-uniform stage segmentation of pipeline parallelism, to evaluate the throughput of HETHUB. The experimental results show that HETHUB has good throughput performance.}

\begin{figure*}[!h]
    \includegraphics[width=170mm]{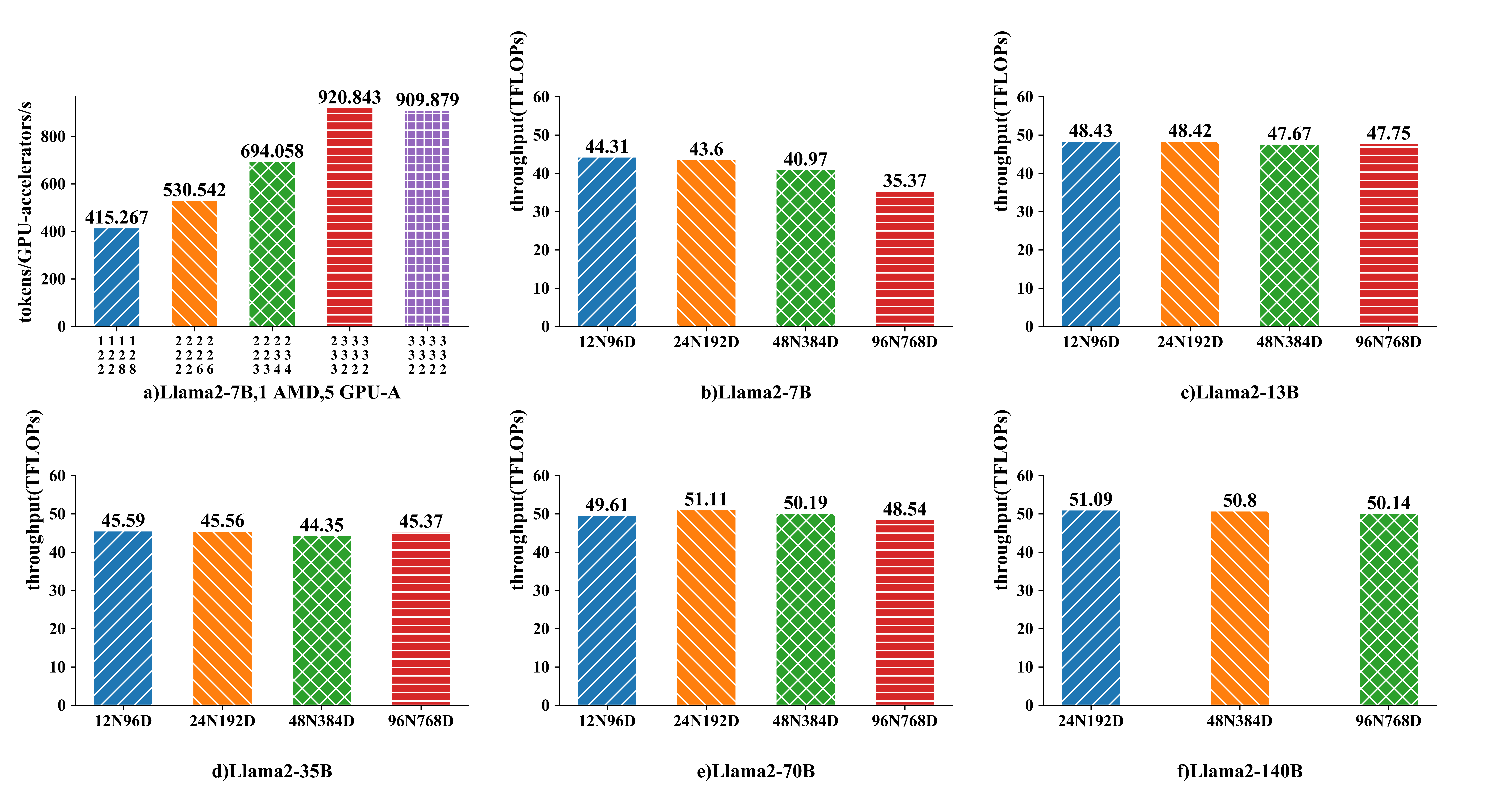}
     \caption{Throughput performance of HETHUB with different models in heterogeneous clusters}
    \label{6_throughput}	
\end{figure*}

\added{The results in Fig.\ref{6_throughput} a) show that the performance of non-uniform stage segmentation is better than that of uniform stage segmentation in heterogeneous clusters. Especially, the un-uniform segmentation with PP = 12 has the highest throughput performance of 920.84 tokens/GPU-accelerators/s, it can improve by up to 2.5$\%$, compared to the uniform segmentation of GPU-accelerator A clusters with PP=6. The experiment results from Fig.\ref{6_throughput}} b)-f) show that the throughput of HETHUB remains stable with the increase of model parameter size and heterogeneous cluster size, and the throughput of HETHUB in heterogeneous clusters reaches 54.71$\%$ of an AMD cluster with 160 GPU-accelerators and 100.96$\%$ for a GPU-accelerator A cluster with 768 GPU-accelerators.  For Llama2-70B, the throughput of AMD is 93.81 TFLOPs/GPU-accelerators, and the throughput of GPU-accelerator A is 48.08 TFLOPs/GPU-accelerators. In the heterogeneous cluster of AMD and More, the throughput reaches a maximum of 51.11 TFLOPs/GPU-accelerators, achieving 91.75$\%$ of the theoretical value for the heterogeneous cluster. However, if the number of cluster nodes increases without changing the model parameter size, the throughput decreases, as the computation per node decreases but cross-node communication increases.

\begin{figure*}[!h]
    \includegraphics[width=170mm]{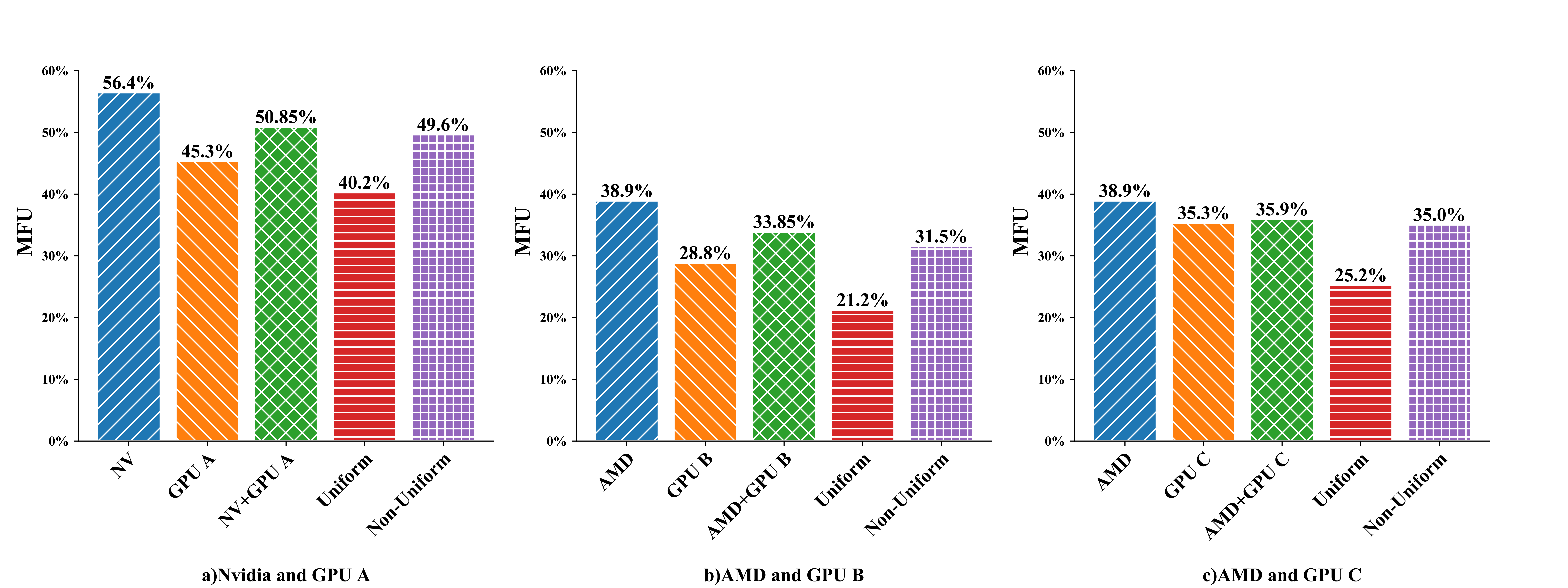}
     \caption{ MFU for training Llama2-70B with uniform and non-uniform stage segmentation in pipeline parallism}
    \label{7_MFU}	
\end{figure*}

\begin{figure*}[!h]
    \includegraphics[width=170mm]{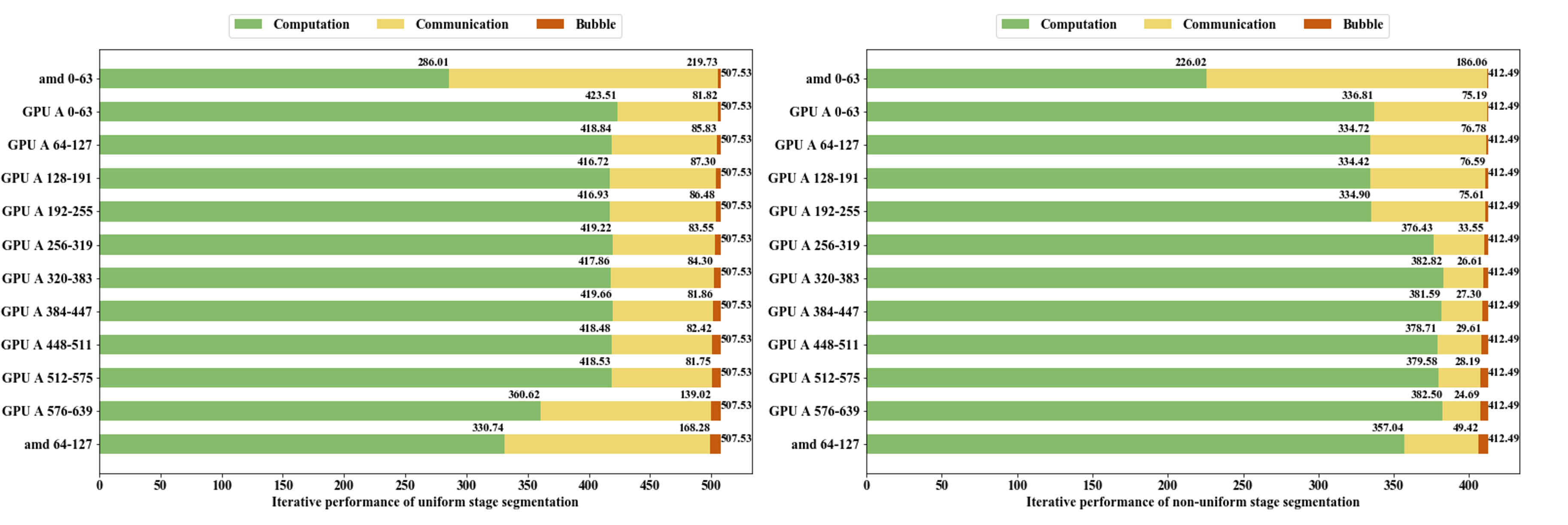}
     \caption{Iterative performance of non-uniform and uniform stage segmentation in pipeline parallism}
    \label{8_Performance}	
\end{figure*}

\subsubsection{MFU}

\added{To evaluate the scalability and computing performance of the HETHUB system further, we tested the MFU of training Llama2-70B model on the homogeneous cluster(Nvidia, AMD, GPU-accelerator A, GPU-accelerator B, and GPU-accelerator C) and heterogeneous cluster(including a 2-node cluster with Nvidia and GPU-accelerator A, a 2-node cluster with AMD and GPU-accelerator B, and a 120-node cluster with AMD and GPU-accelerator C). The experimental results show that the MFU of HETHUB can reach 97.49$\%$ of the theoretical MFU, compared to existing pipeline parallel methods with uniform stage segmentation. Where the PP = 10, a stage has 8 layers for the uniform segmentation method, and PP = 12, stages is 7 6 6 6 6 7 7 7 7 7 7 7, for the Non-uniform stage segmentation.}

\added{Specifically, in Fig.\ref{7_MFU} a), the MFU of Nvidia GPU-accelerator is 56.4\%, the MFU of GPU-accelerator A is 45.3\%, the MFU of heterogeneous clusters with Nvidia and GPU-accelerator A with non-uniform segmentation method is 49.60\%, which can reach 97.54\% of the theoretical MFU 50.85\%. The MFU with the non-uniform segmentation method improves by 9.4\% over the uniform segmentation method. The Fig.\ref{7_MFU} b), the MFU of AMD GPU-accelerator is 38.90\%, the MFU of GPU-accelerator B is 28.80\%, the MFU of heterogeneous clusters with AMD and GPU-accelerator B with non-uniform segmentation method is 31.50\%, which can reach 93.05\% of the theoretical MFU 33.85\%. The MFU with the non-uniform segmentation method improves by 10.3\% over the uniform segmentation method. The Fig.\ref{7_MFU} c), the MFU of AMD GPU-accelerator is 38.90\%, the MFU of GPU-accelerator C is 35.30\%, the MFU of heterogeneous clusters with AMD and GPU-accelerator C with non-uniform segmentation method is 35.00\%, which can reach 97.49\% of the theoretical MFU 35.90\%. The MFU with the non-uniform segmentation method improves by 9.8\% over the uniform segmentation method.}

\added{From the above analysis, we can see that the HETHUB system has excellent scalability,  it can support multiple types of GPU-accleartors for hybrid training large models. Moreover, the non-uniform segmentation method of pipeline parallelism is more suitable for heterogeneous clusters than the uniform segmentation method.}

\subsubsection{End-to-End Performance}
\added{We analyzed the end-to-end performance of the HETHUB system for training the Llama2-70B model on a heterogeneous cluster with 128 AMD GPU-accelerators and 640 GPU-accelerators A, using uniform and non-uniform stage segmentation method of pipeline parallelism. We use pipeline parallelism for models across heterogeneous nodes, AMD, and More. The theoretical bandwidth of Ethernet between the AMD node and the GPU-accelerator A node is 25 Gb/s, the actual bandwidth is 18 - 20 Gb/s in a real environment. The theoretical bandwidth of IB between the AMD node and GPU-accelerator A node is 200 Gb/s, the actual bandwidth is 160 - 180 Gb/s in the real heterogeneous cluster. }

\added{ 
The results in Fig.\ref{8_Performance} show that the end-to-end performance of the non-uniform stage segmentation method of pipeline parallelism is higher than the uniform stage segmentation method, as the differences in the computing resources of different types of GPU-accelerators. The end-to-end time with non-uniform stage segmentation method of pipeline parallelism is 412.49ms, it improves by 18.69\%, compared with the uniform stage segmentation method which needs 507.3ms.}

%% file: conclusion.tex
\section{Conclusion}\label{sec:conc}

\added{In this work, we designed and implemented a distributed training system with hybrid parallelism, HETHUB, for training large-scale models in heterogeneous clusters, including Nvidia, and other types of GPU-accelerators. It supports the communication between different types of GPU-accelerators, and realizes the efficient development, deployment, and training of models through the automatic parallel planner and distributed performance predictor. Experiments demonstrate that the optimal performance of our system in the heterogeneous cluster has achieved up to 97.49$\%$ of the theoretical upper bound performance.}



%% file: main.bbl
\begin{thebibliography}{2}
\providecommand{\natexlab}[1]{#1}
\providecommand{\url}[1]{\texttt{#1}}
\expandafter\ifx\csname urlstyle\endcsname\relax
  \providecommand{\doi}[1]{doi: #1}\else
  \providecommand{\doi}{doi: \begingroup \urlstyle{rm}\Url}\fi

\bibitem[Achiam et~al.(2023)Achiam, Adler, Agarwal, Ahmad, Akkaya, Aleman, Almeida, Altenschmidt, Altman, Anadkat, et~al.]{article_1.0_1}
Josh Achiam, Steven Adler, Sandhini Agarwal, Lama Ahmad, Ilge Akkaya, Florencia~Leoni Aleman, Diogo Almeida, Janko Altenschmidt, Sam Altman, Shyamal Anadkat, et~al.
\newblock Gpt-4 technical report.
\newblock \emph{arXiv preprint arXiv:2303.08774}, 2023.

\bibitem[Zeng et~al.(2021)Zeng, Ren, Su, Wang, Liao, Wang, Jiang, Yang, Wang, Zhang, et~al.]{article_1.0_2}
Wei Zeng, Xiaozhe Ren, Teng Su, Hui Wang, Yi Liao, Zhiwei Wang, Xin Jiang, ZhenZhang Yang, Kaisheng Wang, Xiaoda Zhang, et~al.
\newblock Pangu-$\alpha$: Large-scale autoregressive pretrained chinese language models with auto-parallel computation.
\newblock \emph{arXiv preprint arXiv:2104.12369}, 2021.

\bibitem[Lin et~al.(2021)Lin, Men, Yang, Zhou, Ding, Zhang, Wang, Wang, Jiang, Jia, et~al.]{article_1.0_3}
Junyang Lin, Rui Men, An~Yang, Chang Zhou, Ming Ding, Yichang Zhang, Peng Wang, Ang Wang, Le~Jiang, Xianyan Jia, et~al.
\newblock M6: A chinese multimodal pretrainer.
\newblock \emph{arXiv preprint arXiv:2103.00823}, 2021.

\bibitem[Brown et~al.(2020)Brown, Mann, Ryder, Subbiah, Kaplan, Dhariwal, Neelakantan, Shyam, Sastry, Askell, et~al.]{article_1.0_4}
Tom Brown, Benjamin Mann, Nick Ryder, Melanie Subbiah, Jared~D Kaplan, Prafulla Dhariwal, Arvind Neelakantan, Pranav Shyam, Girish Sastry, Amanda Askell, et~al.
\newblock Language models are few-shot learners.
\newblock \emph{Advances in neural information processing systems}, 33:\penalty0 1877--1901, 2020.

\bibitem[Du et~al.(2021)Du, Qian, Liu, Ding, Qiu, Yang, and Tang]{article_1.0_5}
Zhengxiao Du, Yujie Qian, Xiao Liu, Ming Ding, Jiezhong Qiu, Zhilin Yang, and Jie Tang.
\newblock All nlp tasks are generation tasks: A general pretraining framework, 2021.

\bibitem[Radford et~al.(2019)Radford, Wu, Child, Luan, Amodei, Sutskever, et~al.]{article_1.0_6}
Alec Radford, Jeffrey Wu, Rewon Child, David Luan, Dario Amodei, Ilya Sutskever, et~al.
\newblock Language models are unsupervised multitask learners.
\newblock \emph{OpenAI blog}, 1\penalty0 (8):\penalty0 9, 2019.

\bibitem[Bi et~al.(2023)Bi, Xie, Zhang, Chen, Gu, and Tian]{article_1.0_7}
Kaifeng Bi, Lingxi Xie, Hengheng Zhang, Xin Chen, Xiaotao Gu, and Qi~Tian.
\newblock Accurate medium-range global weather forecasting with 3d neural networks.
\newblock \emph{Nature}, 619\penalty0 (7970):\penalty0 533--538, 2023.

\bibitem[Nori et~al.(2023)Nori, King, McKinney, Carignan, and Horvitz]{article_1.0_8}
Harsha Nori, Nicholas King, Scott~Mayer McKinney, Dean Carignan, and Eric Horvitz.
\newblock Capabilities of gpt-4 on medical challenge problems.
\newblock \emph{arXiv preprint arXiv:2303.13375}, 2023.

\bibitem[Shoeybi et~al.(2019)Shoeybi, Patwary, Puri, LeGresley, Casper, and Catanzaro]{article_1.0_9}
Mohammad Shoeybi, Mostofa Patwary, Raul Puri, Patrick LeGresley, Jared Casper, and Bryan Catanzaro.
\newblock Megatron-lm: Training multi-billion parameter language models using model parallelism.
\newblock \emph{arXiv preprint arXiv:1909.08053}, 2019.

\bibitem[Rasley et~al.(2020)Rasley, Rajbhandari, Ruwase, and He]{article_1.0_10}
Jeff Rasley, Samyam Rajbhandari, Olatunji Ruwase, and Yuxiong He.
\newblock Deepspeed: System optimizations enable training deep learning models with over 100 billion parameters.
\newblock In \emph{Proceedings of the 26th ACM SIGKDD International Conference on Knowledge Discovery \& Data Mining}, pages 3505--3506, 2020.

\bibitem[Li et~al.(2014)Li, Andersen, Park, Smola, Ahmed, Josifovski, Long, Shekita, and Su]{article_1.0_11}
Mu~Li, David~G Andersen, Jun~Woo Park, Alexander~J Smola, Amr Ahmed, Vanja Josifovski, James Long, Eugene~J Shekita, and Bor-Yiing Su.
\newblock Scaling distributed machine learning with the parameter server.
\newblock In \emph{11th USENIX Symposium on operating systems design and implementation (OSDI 14)}, pages 583--598, 2014.

\bibitem[Mnih and Hinton(2008)]{article_1.0_12}
Andriy Mnih and Geoffrey~E Hinton.
\newblock A scalable hierarchical distributed language model.
\newblock \emph{Advances in neural information processing systems}, 21, 2008.

\bibitem[Luo et~al.(2020)Luo, He, Zhuo, and Qian]{article_1.0_13}
Qinyi Luo, Jiaao He, Youwei Zhuo, and Xuehai Qian.
\newblock Prague: High-performance heterogeneity-aware asynchronous decentralized training.
\newblock In \emph{Proceedings of the Twenty-Fifth International Conference on Architectural Support for Programming Languages and Operating Systems}, pages 401--416, 2020.

\bibitem[Sergeev and Balso(2018)]{2018Horovod}
Alexander Sergeev and Mike~Del Balso.
\newblock Horovod: fast and easy distributed deep learning in tensorflow.
\newblock \emph{CoRR}, abs/1802.05799, 2018.
\newblock URL \url{http://arxiv.org/abs/1802.05799}.

\bibitem[Li et~al.(2020)Li, Zhao, Varma, Salpekar, and Chintala]{2020PyTorch}
Shen Li, Yanli Zhao, Rohan Varma, Omkar Salpekar, and Soumith Chintala.
\newblock Pytorch distributed: Experiences on accelerating data parallel training.
\newblock 2020.

\bibitem[Duchi et~al.(2011)Duchi, Hazan, and Singer]{2011Adaptive}
John Duchi, Elad Hazan, and Yoram Singer.
\newblock Adaptive subgradient methods for online learning and stochastic optimization.
\newblock pages 257--269, 2011.


\bibitem[Rajbhandari et~al.(2019)Rajbhandari, Rasley, Ruwase, and He]{DBLP:journals/corr/abs-1910-02054}
Samyam Rajbhandari, Jeff Rasley, Olatunji Ruwase, and Yuxiong He.
\newblock Zero: Memory optimization towards training {A} trillion parameter models.
\newblock \emph{CoRR}, abs/1910.02054, 2019.
\newblock URL \url{http://arxiv.org/abs/1910.02054}.

\bibitem[Huang et~al.(2019)Huang, Cheng, Bapna, Firat, Chen, Chen, Lee, Ngiam, Le, Wu, et~al.]{article_2.0_1}
Yanping Huang, Youlong Cheng, Ankur Bapna, Orhan Firat, Dehao Chen, Mia Chen, HyoukJoong Lee, Jiquan Ngiam, Quoc~V Le, Yonghui Wu, et~al.
\newblock Gpipe: Efficient training of giant neural networks using pipeline parallelism.
\newblock \emph{Advances in neural information processing systems}, 32, 2019.

\bibitem[Harlap et~al.(2018)Harlap, Narayanan, Phanishayee, Seshadri, Devanur, Ganger, and Gibbons]{pipedream-1f1b}
Aaron Harlap, Deepak Narayanan, Amar Phanishayee, Vivek Seshadri, Nikhil Devanur, Greg Ganger, and Phil Gibbons.
\newblock Pipedream: Fast and efficient pipeline parallel dnn training.
\newblock \emph{arXiv preprint arXiv:1806.03377}, 2018.

\bibitem[Narayanan et~al.(2021)Narayanan, Phanishayee, Shi, Chen, and Zaharia]{pipedream-2bw}
Deepak Narayanan, Amar Phanishayee, Kaiyu Shi, Xie Chen, and Matei Zaharia.
\newblock Memory-efficient pipeline-parallel dnn training.
\newblock In \emph{International Conference on Machine Learning}, pages 7937--7947. PMLR, 2021.

\bibitem[Li and Hoefler(2021)]{article_2.0_3}
Shigang Li and Torsten Hoefler.
\newblock Chimera: efficiently training large-scale neural networks with bidirectional pipelines.
\newblock In \emph{Proceedings of the International Conference for High Performance Computing, Networking, Storage and Analysis}, pages 1--14, 2021.

\bibitem[Fan et~al.(2021)Fan, Rong, Meng, Cao, Wang, Zheng, Wu, Long, Yang, Xia, et~al.]{fan2021dapple}
Shiqing Fan, Yi~Rong, Chen Meng, Zongyan Cao, Siyu Wang, Zhen Zheng, Chuan Wu, Guoping Long, Jun Yang, Lixue Xia, et~al.
\newblock Dapple: A pipelined data parallel approach for training large models.
\newblock In \emph{Proceedings of the 26th ACM SIGPLAN Symposium on Principles and Practice of Parallel Programming}, pages 431--445, 2021.

\bibitem[Santhanam et~al.(2021)Santhanam, Krishna, Tomioka, Fitzgibbon, and Harris]{distir}
Keshav Santhanam, Siddharth Krishna, Ryota Tomioka, Andrew Fitzgibbon, and Tim Harris.
\newblock Distir: An intermediate representation for optimizing distributed neural networks.
\newblock In \emph{Proceedings of the 1st Workshop on Machine Learning and Systems}, pages 15--23, 2021.

\bibitem[Schaarschmidt et~al.(2021)Schaarschmidt, Grewe, Vytiniotis, Paszke, Schmid, Norman, Molloy, Godwin, Rink, Nair, et~al.]{automap}
Michael Schaarschmidt, Dominik Grewe, Dimitrios Vytiniotis, Adam Paszke, Georg~Stefan Schmid, Tamara Norman, James Molloy, Jonathan Godwin, Norman~Alexander Rink, Vinod Nair, et~al.
\newblock Automap: Towards ergonomic automated parallelism for ml models.
\newblock \emph{arXiv preprint arXiv:2112.02958}, 2021.

\bibitem[Jia et~al.(2019)Jia, Zaharia, and Aiken]{jia2019beyond}
Zhihao Jia, Matei Zaharia, and Alex Aiken.
\newblock Beyond data and model parallelism for deep neural networks.
\newblock \emph{Proceedings of Machine Learning and Systems}, 1:\penalty0 1--13, 2019.

\bibitem[Wang et~al.(2021)Wang, Li, Tachon, Wang, Yang, Limet, and Robert]{article_2.0_5}
Haoran Wang, Chong Li, Thibaut Tachon, Hongxing Wang, Sheng Yang, S{\'e}bastien Limet, and Sophie Robert.
\newblock Efficient and systematic partitioning of large and deep neural networks for parallelization.
\newblock In \emph{Euro-Par 2021: Parallel Processing: 27th International Conference on Parallel and Distributed Computing, Lisbon, Portugal, September 1--3, 2021, Proceedings 27}, pages 201--216. Springer, 2021.

\bibitem[Tarnawski et~al.(2021)Tarnawski, Narayanan, and Phanishayee]{piper}
Jakub~M Tarnawski, Deepak Narayanan, and Amar Phanishayee.
\newblock Piper: Multidimensional planner for dnn parallelization.
\newblock \emph{Advances in Neural Information Processing Systems}, 34:\penalty0 24829--24840, 2021.

\bibitem[Zheng et~al.(2022)Zheng, Li, Zhang, Zhuang, Chen, Huang, Wang, Xu, Zhuo, Xing, et~al.]{article_2.0_6}
Lianmin Zheng, Zhuohan Li, Hao Zhang, Yonghao Zhuang, Zhifeng Chen, Yanping Huang, Yida Wang, Yuanzhong Xu, Danyang Zhuo, Eric~P Xing, et~al.
\newblock Alpa: Automating inter-and $\{$Intra-Operator$\}$ parallelism for distributed deep learning.
\newblock In \emph{16th USENIX Symposium on Operating Systems Design and Implementation (OSDI 22)}, pages 559--578, 2022.


\bibitem[Chen et~al.(2018)Chen, Li, Bilal, Li, Philip, et~al.]{bpt_cnn_2018}
Jianguo Chen, Kenli Li, Kashif Bilal, Keqin Li, S~Yu Philip, et~al.
\newblock A bi-layered parallel training architecture for large-scale convolutional neural networks.
\newblock \emph{IEEE transactions on parallel and distributed systems}, 30\penalty0 (5):\penalty0 965--976, 2018.

\bibitem[Song et~al.(2020)Song, Chen, Zhuo, Qian, Li, and Chen]{accpar_2020}
Linghao Song, Fan Chen, Youwei Zhuo, Xuehai Qian, Hai Li, and Yiran Chen.
\newblock Accpar: Tensor partitioning for heterogeneous deep learning accelerators.
\newblock In \emph{2020 IEEE International Symposium on High Performance Computer Architecture (HPCA)}, pages 342--355. IEEE, 2020.

\bibitem[Jia et~al.(2022)Jia, Jiang, Wang, Xiao, Shi, Zhang, Li, Chen, Li, Zheng, et~al.]{whale_2022}
Xianyan Jia, Le~Jiang, Ang Wang, Wencong Xiao, Ziji Shi, Jie Zhang, Xinyuan Li, Langshi Chen, Yong Li, Zhen Zheng, et~al.
\newblock Whale: Efficient giant model training over heterogeneous $\{$GPUs$\}$.
\newblock In \emph{2022 USENIX Annual Technical Conference (USENIX ATC 22)}, pages 673--688, 2022.

\bibitem[Smith et~al.(2022)Smith, Patwary, Norick, LeGresley, Rajbhandari, Casper, Liu, Prabhumoye, Zerveas, Korthikanti, et~al.]{article_3.0_2}
Shaden Smith, Mostofa Patwary, Brandon Norick, Patrick LeGresley, Samyam Rajbhandari, Jared Casper, Zhun Liu, Shrimai Prabhumoye, George Zerveas, Vijay Korthikanti, et~al.
\newblock Using deepspeed and megatron to train megatron-turing nlg 530b, a large-scale generative language model.
\newblock \emph{arXiv preprint arXiv:2201.11990}, 2022.

\bibitem[Facebook(2023)]{article_3.0_1}
Gloo 2023.
\newblock Collective Communications Library with Various Primitives for Multi-Machine Training. 
\newblock \emph{https://github.com/facebookincubator/gloo}, 2023

\bibitem[Kashyap(2006)]{article_3.0_8}
Vivek Kashyap.
\newblock Ip over infiniband (ipoib) architecture.
\newblock RFC 4392, April 2006.
\newblock URL \url{https://www.rfc-editor.org/info/rfc4392}.

\bibitem[Kalia et~al.(2014)Kalia, Kaminsky, and Andersen]{article_3.0_9}
Anuj Kalia, Michael Kaminsky, and David~G. Andersen.
\newblock Using rdma efficiently for key-value services.
\newblock \emph{ACM SIGCOMM Computer Communication Review}, 44\penalty0 (4):\penalty0 295--306, 2014.

\bibitem[Kennedy and Kremer(1998)]{3_np_hard}
Ken Kennedy and Ulrich Kremer.
\newblock Automatic data layout for distributed-memory machines.
\newblock \emph{ACM Transactions on Programming Languages and Systems (TOPLAS)}, 20\penalty0 (4):\penalty0 869--916, 1998.

\bibitem[Liang et~al.(2023)Liang, Tang, Zhang, Bai, Su, Lai, Qiao, and Li]{3_auto_survey}
Peng Liang, Yu~Tang, Xiaoda Zhang, Youhui Bai, Teng Su, Zhiquan Lai, Linbo Qiao, and Dongsheng Li.
\newblock A survey on auto-parallelism of large-scale deep learning training.
\newblock \emph{IEEE Transactions on Parallel and Distributed Systems}, 2023.

\bibitem[Touvron et~al.(2023)Touvron, Martin, Stone, Albert, Almahairi, Babaei, Bashlykov, Batra, Bhargava, Bhosale, et~al.]{article_4.0_1}
Hugo Touvron, Louis Martin, Kevin Stone, Peter Albert, Amjad Almahairi, Yasmine Babaei, Nikolay Bashlykov, Soumya Batra, Prajjwal Bhargava, Shruti Bhosale, et~al.
\newblock Llama 2: Open foundation and fine-tuned chat models, 2023.

\bibitem[Soldaini et~al.(2024)Soldaini, Kinney, Bhagia, Schwenk, Atkinson, Authur, Bogin, Chandu, Dumas, Elazar, Hofmann, Jha, Kumar, Lucy, Lyu, Lambert, Magnusson, Morrison, Muennighoff, Naik, Nam, Peters, Ravichander, Richardson, Shen, Strubell, Subramani, Tafjord, Walsh, Zettlemoyer, Smith, Hajishirzi, Beltagy, Groeneveld, Dodge, and Lo]{soldaini2024dolma}
Luca Soldaini, Rodney Kinney, Akshita Bhagia, Dustin Schwenk, David Atkinson, Russell Authur, Ben Bogin, Khyathi Chandu, Jennifer Dumas, Yanai Elazar, Valentin Hofmann, Ananya~Harsh Jha, Sachin Kumar, Li~Lucy, Xinxi Lyu, Nathan Lambert, Ian Magnusson, Jacob Morrison, Niklas Muennighoff, Aakanksha Naik, Crystal Nam, Matthew~E. Peters, Abhilasha Ravichander, Kyle Richardson, Zejiang Shen, Emma Strubell, Nishant Subramani, Oyvind Tafjord, Pete Walsh, Luke Zettlemoyer, Noah~A. Smith, Hannaneh Hajishirzi, Iz~Beltagy, Dirk Groeneveld, Jesse Dodge, and Kyle Lo.
\newblock Dolma: an open corpus of three trillion tokens for language model pretraining research, 2024.


\bibitem[Du et~al.(2024)Du, Yu, Gao, Pan, Cheng, Ma, Yuan, Qu, Liu, Zheng, Luo, Zhou, Yuan, Chen, Fu, and Zhang]{mapcc_2024}
Xinrun Du, Zhouliang Yu, Songyang Gao, Ding Pan, Yuyang Cheng, Ziyang Ma, Ruibin Yuan, Xingwei Qu, Jiaheng Liu, Tianyu Zheng, Xinchen Luo, Guorui Zhou, Binhang Yuan, Wenhu Chen, Jie Fu, and Ge~Zhang.
\newblock Chinese tiny llm: Pretraining a chinese-centric large language model, 2024.



























































\bibitem[Zhang et~al.(2021)Zhang, Yin, Wu, and Zhou]{article_3.0_6}
Rui Zhang, Zhendong Yin, Zhilu Wu, and Siyang Zhou.
\newblock A novel automatic modulation classification method using attention mechanism and hybrid parallel neural network.
\newblock \emph{Applied Sciences}, 11\penalty0 (3):\penalty0 1327, 2021.





\end{thebibliography}
